\documentclass[oribibl]{llncs}

\usepackage{amssymb}	 
\usepackage{amsmath}

\usepackage[colorlinks=true, linkcolor=blue, citecolor=blue, urlcolor=blue]{hyperref}

\usepackage{graphicx}
\usepackage{booktabs}
\usepackage{multirow}
\usepackage{tabularx}
\usepackage{threeparttable}
\usepackage{url}

\usepackage{fancyhdr} 
\fancypagestyle{firststyle}
{
   \fancyhf{}
   \fancyfoot[C]{\scriptsize{Proceedings of the 23rd IFIP Conference on e-Business, e-Services, and e-Society (I3E 2024), Lecture Notes in Computer Science (vol.~14907), Heerlen, Springer, pp. 284–295. This version is the author's copy. The publisher's definite version is available at Springer via \url{https://doi.org/10.1007/978-3-031-72234-9_24}}}
}

\begin{document}

\title{The Incoherency Risk in the EU's \\ New Cyber Security Policies}

\author{Jukka Ruohonen\orcidID{0000-0001-5147-3084} \email{juk@mmmi.sdu.dk}}
\institute{University of Southern Denmark, S\o{}nderborg, Denmark}

\maketitle

\begin{abstract}
The European Union (EU) has been pursuing new cyber security policies in recent
years. This paper presents a short reflection of four such policies. The focus
is on potential incoherency, meaning a lack of integration, divergence between
the member states, institutional dysfunction, and other related problems that
should be at least partially avoidable by sound policy-making. According to the
results, the four policies have substantially increased the complexity of the
EU's cyber security framework. In addition, there are potential problems with
trust, divergence between industry sectors and different technologies,
bureaucratic conflicts, and technical issues, among other things. With these
insights, the paper not only contributes to the study of EU policies but also
advances the understanding of cyber security policies in general.
\end{abstract}

\begin{keywords}
cyber security, policy coherency, CER, NIS2, CSA, CRA, EU
\end{keywords}

\section{Introduction}

\thispagestyle{firststyle} 

The historical origins of the EU's framework for cyber security trace to the
late 1990s and early 2000s during which the first policies and institutional
arrangements were established~\cite{Ruohonen16GIQ}. Since then, the framework
has been continuously altered and updated as cyber security has gained
prominence---and vicious cyber attacks have continued. Throughout the EU, these
attacks, whether they manifest themselves as data breaches, ransomware
operations, or industrial espionage, cause substantial financial losses,
reputation damage, and even human misery. Geopolitical conflicts and tensions,
disruptive new technologies such as artificial intelligence~(AI), risks related
to critical infrastructure and even democracy, and related major factors have
further increased the importance of cyber security. To this end, the European
Commission published in 2020 a new strategy for cyber security in the
EU. According to this strategy, improving cyber security is necessary for
enjoying the benefits from innovation, connectivity, and automation, while still
maintaining fundamental rights and freedoms~\cite{EC20e}. Cyber security is thus
an enabler on one hand. On the other, it is also a source for innovations in
itself. While regulations are often seen as a hindrance to industry and its
competitiveness, it is therefore important to consider also policy innovations
that may improve not only cyber security but also industry competitiveness.

For a long time, there was a lack of evaluation studies on the various digital
policies enacted throughout the world~\cite{Leeuw12}. While the situation has
improved over time, critical evaluations and reflections are always needed as
these allow to also understand the potential challenges ahead. With this point
in mind, the paper at hand presents a brief critical reflection on four recent
cyber security policies, some of which are already enacted EU laws. All belong
to the noted cyber security strategy pursued by the von der Leyen's first
Commission. Furthermore, the other traditional pillars of the EU's cyber
security framework are mostly omitted; these pillars are cyber crime, cyber
defense, and, to a lesser extent, data protection~\cite{Christou16}. The
critical reflection pursued is motivated by potential policy incoherency. This
concept is elaborated in the opening Section~\ref{sec: coherency}. After this
elaboration, the four policies are briefly discussed in the subsequent
Section~\ref{sec: policies}. Then, the potential incoherencies arising from the
four new cyber security policies are elaborated and speculated in
Section~\ref{sec: incoherencies}. The last Section~\ref{sec: conclusion}
concludes. Finally, it should be emphasized that the paper is a reflection on a
complex policy constellation. Actual evaluation can only be done in a distant
future once the policies have been adopted by the member states and the
practical operation and administration of these have continued for a sufficient
time. This point justifies the term risk in the paper's title; the focus is on
potential and not (yet) actual problems under uncertainty about the
future. Methodologically the critical reflection falls to the domain of
interpretive policy analysis~\cite{Munch16, Yanow00}. That is, the
methodological approach is not an ``objective'' cost-benefit analysis or other
related mechanical evaluation, but rather a short critical interpretation of the
policies and their impact under the objective to improve cyber~security.

\section{Policy Coherency and Incoherency}\label{sec: coherency}

The EU has increased the interdependence of its member states and thus policy
convergence has been an important goal particularly with respect to the internal
market. In short: the rules should be the same across all member states. To
achieve convergence, a policy needs to be coherent. A coherent policy is
well-integrated, meaning that its diverse components work well
together~\cite{Bolognesi19}. Lack of integration has been a typical problem in
the cyber security domain. In many countries different cyber security agencies
and actors have formed their own silos beyond which cooperation may not work and
critical information does not necessarily traverse~\cite{Hurel21}. The EU is not
an exception; different EU agencies and also the member states themselves have
often formed their own silos~\cite{CEPS18, Christou16}. While a EU policy
typically contains many internal components, integration should thus apply also
across the member states and through EU-level administration.

It is useful to distinguish different forms of policy coherency. Thus, internal
coherency can be understood to refer to the consistency between the goals,
objectives, modalities, and protocols of a particular EU policy; intra-country
coherency can be seen to involve the consistency of a member state's policies in
terms of their combined contribution to the EU; and, finally, inter-country
coherency can be interpreted as the consistency and uniformity of policies
across several member states with respect to their aggregated impact at the
EU-level~\cite{Picciotto05}. Extensive transaction costs are a typical sign of
problems in internal coherency; these costs are due to a lack of information, a
need for monitoring and control, and enforcement measures and administrative
support~\cite{Bolognesi19}. Ideally, such costs can be decreased by good
planning. The actors involved also learn over time, which should also eventually
reduce the transaction costs.  Analytically, in the EU context, it is useful to
further think in terms of vertical and horizontal policy
coherency~\cite{Portela12, Ruohonen22ICLR}. Vertical coherency refers to the
alignment between the EU and the member states; the aggregated impact to the EU
in intra-country and inter-country coherency is thus vertical. In contrast,
horizontal policy coherency refers to the uniformity of either EU-level
arrangements or the member states' policies and the convergence between
them. These policies can be purely national laws or transposed EU laws. This
simple analytical machinery for policy coherency allows to also approach the
antonym, policy incoherency.

A traditional form of policy incoherency in the EU context is vertical; a
mismatch between EU laws and their national adaptations. In purely judicial
cases, there are well-established mechanisms for dealing with this kind of
incoherency; the European Commission may even start an infringement proceedings
against a member state for incorrect national transposition of a given EU
law. Another form of vertical policy incoherency may arise when the roles and
responsibilities are not defined clearly enough between EU-level and national
institutions and institutional practices. A coherent policy solution requires
also sound alignment between institutions and
technologies~\cite{Perennes13}. For instance, national information exchanges
toward EU-level institutions may work poorly due to ambiguous legal guidelines
or poor technical specifications, both of which are likely to increase
transaction costs. A~potential dysfunction may further affect coordination more
generally. With respect to coordination, particularly important at the EU-level
is the European Union Agency for Cybersecurity (ENISA), although there are many
other relevant actors as well. A risk of either vertical or horizontal
incoherency is further solidified by the recent policies that have further
increased the number of cyber security actors both nationally and at the
EU-level.

Also horizontal policy incoherencies have been common in the EU. A typical form
of horizontal (inter-country) incoherency is a mismatch between national
adaptations of a EU law. In other words, a particular EU law may differ from a
member state to another due to different national interpretations and
adaptations. Countering this type of horizontal incoherency with harmonization
has been a long-standing goal in the EU. Further institutional horizontal
incoherencies may arise both within national public administrations and EU-level
administration. In terms of the former, ambiguities may cause bureaucratic
conflicts between different national public administration units. Data
protection would be a good example because national data protection authorities
are typically distinct from cyber security authorities, which are distinct from
law enforcement authorities, although all may be required to address some cyber
security incidents such as data breaches. Analogous conflicts may emerge between
different industry sectors. Such conflicts are non-optimal because a coherent
policy should be consistent, comprehensive, and harmonious across domains and
sectors in a manner that does not compromise its goals~\cite{Dube14}. Even
though operational cyber security has thus far been limited at the EU-level,
somewhat analogous institutional conflicts may still emerge also at the EU-level
administration.

Finally, a policy incoherency may become more serious if the misguided policy
outcomes become permanent. In the present context, a particular concern is
whether a policy incoherency may turn into a permanent institutional
incoherency. Unlike policies, which, at least in theory, may be altered and
updated relatively easily, institutions are much more salient and permanent. In
fact, this institutional stickiness is one of the primary takeaways from the
extensive path-dependency literature~\cite{Mahoney00, Ruohonen16GIQ}. In the EU
context these institutions can be understood to refer to permanent arrangements
or agencies with recruited staff who typically work in fixed physical
locations. While there is room for institutional innovation when designing new
EU policies, also the incoherency risks from poor institutional planning are
thus clear. It is not easy to backpedal once a specific agency or a permanent
bureaucratic arrangement has been established. The same point applies to
technologies and technical solutions, especially since some of the new policies
envision large pan-European cyber security solutions.

\section{The Policies in Brief}\label{sec: policies}

\begin{figure*}[th!b]
\centering
\includegraphics[width=\linewidth, height=6.8cm]{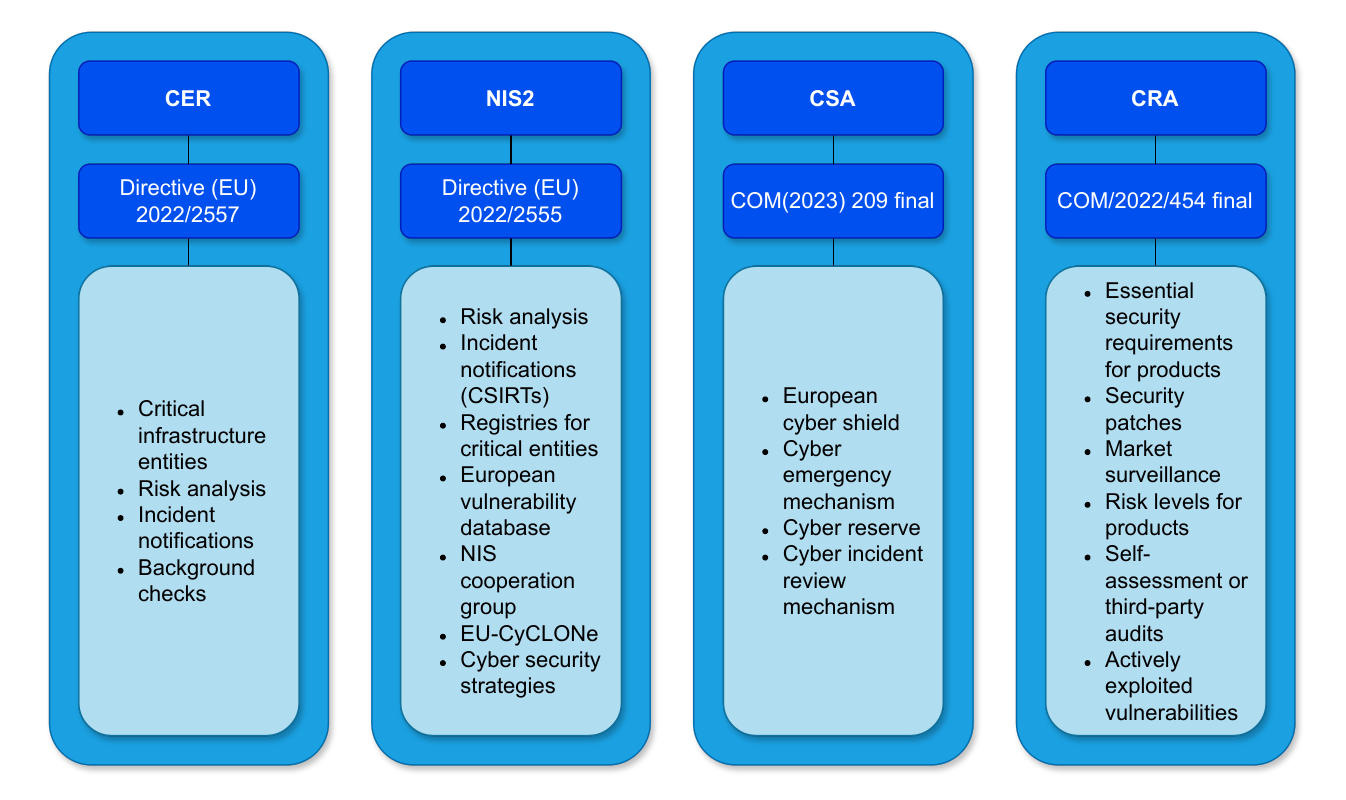}
\caption{The Four Cyber Security Policies}
\label{fig: policies}
\end{figure*}

There are four cyber security policies that are important for approaching
potential policy incoherencies in the EU. These are illustrated in
Fig.~\ref{fig: policies}. Two of these are already enacted laws, while two are
still under political decision-making.

The first law is the Critical Entities Resilience (CER) directive. This
directive substantially increases the amount of critical infrastructure entities
from an older directive that only considered the energy and transportation
sectors. Now, also the banking, financial market, healthcare, water, and food
sectors together with core elements of the Internet's infrastructure, public
administration, and space technologies are defined as critical. Regarding the
Internet's infrastructure, domain name system services, top-level domain name
registries, data centers, content delivery networks, and cloud computing
services are covered, among a few other technologies. While leeway is left for
the member states regarding the details on how the critical entities are
identified, the CER directive imposes also many requirements for the identified
critical entities. Among these are risk assessments for natural and man-made
risks, incident notifications for significant disruptions, and background checks
for employees operating the critical entities.

The second law is the so-called NIS2 directive. Based on a significant update of
the older NIS directive, it essentially augments the requirements from the CER
directive for the critical entities. Particularly risk assessments are specified
in more detail; these should cover not only traditional information security and
incident handling, but also business continuity, supply chain security, security
of equipment acquisitions, employee training, use of cryptography, and
organizational security should be assessed. On the public sector side, the
member states are obliged to designate one or more competent authorities with a
single point of contact for incident reporting. Substantial enforcement powers
are also granted for the competent authorities; administrative fines are part of
the enforcement toolbox. At the EU-level, ENISA is tasked to administer the
registries for the critical Internet technologies and to build a European
vulnerability database. As for incidents, as before, these are reported to
national computer security incident response teams (CSIRTs) whose obligations
are also extended, including with respect to coordinated vulnerability
disclosure and cooperation at the EU-level. In addition to the management of
large-scale cyber security incidents and the national enforcement of the
directive, the competent authorities have a liaison function for cross-border
investigations conducted together with the Commission and ENISA. Furthermore,
the NIS2 directive specifies a NIS cooperation group and European cyber crisis
liaison organisation network (EU-CyCLONe) for improved EU-level coordination,
including with respect to information exchanges and situational
awareness. Finally, the NIS2 directive obliges each member state to publish a
national cyber security strategy.

The Cyber Solidarity Act (CSA) law proposal is the third relevant cyber security
policy to consider. The proposal's motivation originates from serious
cross-border cyber crises in the face of the war in Ukraine. The objectives of
the proposal are to be implemented with pan-European security operations centers
(SOCs) together with a cyber emergency mechanism and a review mechanism for
large cyber security incidents. In practice, the SOCs discussed amount to large,
either national or cross-border, threat intelligence platforms; the goal is to
enhance attack detection, prevention, and response. Through EU-based funding
available via the Digital Europe Programme (DEP), national SOCs are expected to
enlarge into cross-border SOCs, eventually creating a unified ``European cyber
shield''. Information obtained through cross-border SOCs about large-scale cyber
security incidents should be shared between the CSIRT network, EU-CyCLONe, and
the Commission. The cyber emergency mechanism, in turn, is tailored for
preparedness actions, recovery actions from large-scale incidents, and mutual
assistance between the member states. Of these, the mutual assistance provision
is accompanied with a specific cyber security reserve that can be deployed to
any given member state upon a request to the Commission and ENISA, and informing
of the CSIRT network and where appropriate the EU-CyCLONe network. In practice,
the reserve is composed of members from the CSIRT network together with
personnel from EU agencies, although support from like-minded third countries is
also allowed. Finally, the incident review mechanism is planned for \textit{post
  hoc} analysis and review of a large-scale cyber security incident. Most of the
work involved is delegated to ENISA.

The fourth and final law proposal is the Cyber Resilience Act (CRA). It is based
on a different rationale than the three other policies. The proposal is
motivated by consumer protection jurisprudence, the low level of cyber security
in many information technology products, as reflected by widespread
vulnerabilities in them and inconsistent provision of security patches to
counter the vulnerabilities, and insufficient information for consumers to make
informed choices regarding cyber security. To this end, the proposal obliges
manufacturers to take security into account by specifying essential security
requirements, including for coordinated vulnerability disclosure. In addition,
manufacturers are mandated to provide security patches for the whole life cycle
of their products. On the public sector side, the proposal lays down rules on
market surveillance and enforcement. Administrative fines are again part of the
enforcement arsenal. Although a few specialized products are excluded, the
proposal in practice covers the whole information technology sector. Also
software products are in the scope. Furthermore, the proposal categorizes
products according to their cyber security risk level. While the manufacturing
of most products is possible under self-assessment of conformity, products
categorized as critical require third-party audits. Regarding coordinated
vulnerability disclosure, manufacturers are further obliged to inform ENISA
about any actively exploited vulnerabilities in their products. According to the
proposal, ENISA will then inform the national CSIRTs.

Finally, it should be emphasized that the CSA and CRA proposals are still being
negotiated. Therefore, the final laws may differ from the original law proposals
considered. All policies except the CSA proposal also grant the Commission a
power to adopt delegated acts to supplement the policies in different ways. In
general, these acts increase the Commission's power to further shape the EU's
new cyber security policy framework.

\section{Potential Incoherencies}\label{sec: incoherencies}

The first potential horizontal policy incoherency is the traditional one, the
divergence between national transpositions of the EU laws in the member
states. In particular, CER and NIS2 are directives, which give more room for
national transpositions than regulations, including the CSA and CRA law
proposals, which are directly applicable EU law.

The second potential incoherency originates from the background checks specified
in the CER directive. The type of potential incoherency is horizontal. Namely,
there may be some conflicts in national adaptations of the directive because
background checks are typically done by law enforcement or intelligence
agencies, whereas, as per the NIS2 directive, incident notifications are
typically delivered to different organizations in most member
states. Furthermore, a traditional horizontal incoherency between the member
states is likely because there is no harmonization in the EU for background
checks or security clearances~\cite{Ruohonen20EJSR}. Because some of the
critical entities are operated by multinational companies, it may also be
difficult to carry out background checks in practice as there are no
extraterritorial powers for such checks.

The third potential horizontal incoherency is also related to the CER and NIS2
directives. Namely, the critical entities specified in the directives should be
synchronized for sound administration and enforcement. Due to the previous
point, some incoherency may still appear. For instance, it may be that
background checks are not done uniformly across all critical entities in some
member states. If synchronization is lacking, duplication of work may also occur
with the risk assessments specified in both directives.

The fourth potential incoherency continues the same horizontal theme; the
policies put different industry sectors and technologies into a uneven
position. This potential incoherency is present with respect to the critical
entities specified in the CER and NIS2 directives; the non-critical sectors left
behind may receive less attention and resources for cyber
security. Alternatively, it can be argued that the directives specify too many
sectors as critical, which may cause different interpretation problems and lead
to incoherent implementation and inconsistent enforcement across the member
states~\cite{Mikac23}. Analogous consequences are present with respect to the
CRA proposal; the products defined as non-critical presumably receive much less
scrutiny than the critical products. This uneven treatment of sectors and
technologies carries also economic consequences. Given the divergence of the
information technology sectors in the member states, the uneven treatment may
even lead to a larger horizontal incoherency across the union.

The fifth potential policy incoherency is related to the additional
administrative bodies specified in the NIS2 directive and the CSA proposal. In
particular, the new EU-CyCLONe network may cause redundancy because the CSIRT
network is already well-established in Europe. In fact, the EU's cyber security
policy framework was from the beginning built upon the CSIRT network with ENISA
acting in a coordination role at the EU-level~\cite{Ruohonen16GIQ}. Analogous
redundancy issues may arise with respect to the cyber reserve specified in the
CSA proposal. Many problematic scenarios are possible due to the redundancy and
related duplication. For instance, horizontal incoherency may again manifest
itself through bureaucratic conflicts, starting from the recruitment of staff to
the EU-CyCLONe network, the CSIRT network, the NIS cooperation group, and the
cyber reserve.

Trust is behind the sixth potential policy incoherency, which may be both
horizontal and vertical. Because decision-making is centralized to the EU-level,
there may be potential issues in the deployment of the cyber reserve envisioned
in the CSA proposal in case a member state in need lacks trust toward ENISA or
the Commission. Alternatively, there is no strict deadlines for the deployment
decision-making; therefore, as many cyber security issues require urgent
attention, the deployment may become moot especially under multiple and
concurrent requests~\cite{ECA24}. Furthermore, the allowance of third countries
in the reserve may prevent a member state from requesting help in case it lacks
trust toward a given third country. While it is difficult to evaluate how
realistic these scenarios are, a bigger potential incoherency is also present;
it remains unclear how the cyber reserve will align with similar initiatives
taken under the North Atlantic Treaty Organization~(NATO). Duplication of effort
and resources is a real possibility. There is also a more deep-rooted question
about civilian and military approaches to cyber
security~\cite{Ruohonen20EJSR}. Particularly the cyber reserve may blur the
boundary between the two. Such blurring of fundamental boundaries might perhaps
even lead to a philosophical incoherency in the EU's cyber security framework.

The seventh set of potential incoherencies is related to the SOCs specified in
the CSA proposal. A potential vertical incoherency may be present because the
proposal does not say much about how EU-level agencies should share data with
national SOCs~\cite{ECA24}. Furthermore, as cross-border SOCs are expected to
share information about large-scale incidents rather widely across different
actors, hypothetical conflicts and trust issues may potentially spill also to
the domain of operational security. Because SOCs are about operational security,
these presumably process and transmit sensitive data, which may have even
national security consequences. Therefore, lack of trust between the member
states may cause difficulties for the establishment of cross-border SOCs to
begin with. It is difficult to say whether the financial carrots from the DEP
can fully resolve this trust conundrum. Finally, there are significant technical
interoperability challenges in building large cross-border SOCs and integrating
these to already existing national SOCs. The CSA proposal also eagerly speaks
about using AI for the threat intelligence available through SOCs. However,
during the negotiations, some skepticism was expressed about whether AI
technologies are mature enough to be used in the envisioned European cyber
shield. While it is too early to say anything about actual implementations and
their funding through the DEP or other instruments, it can be still concluded
that also technologies and technical issues may contribute to the potential
incoherency. Similar points apply with respect to SOCs already established in
the private sector. While these were to some extent addressed during the
negotiations~\cite{Clasen23}, the original proposal did not consider cooperation
and information exchanges with private sector actors.

\begin{figure*}[th!b]
\centering
\includegraphics[width=\linewidth, height=12.2cm]{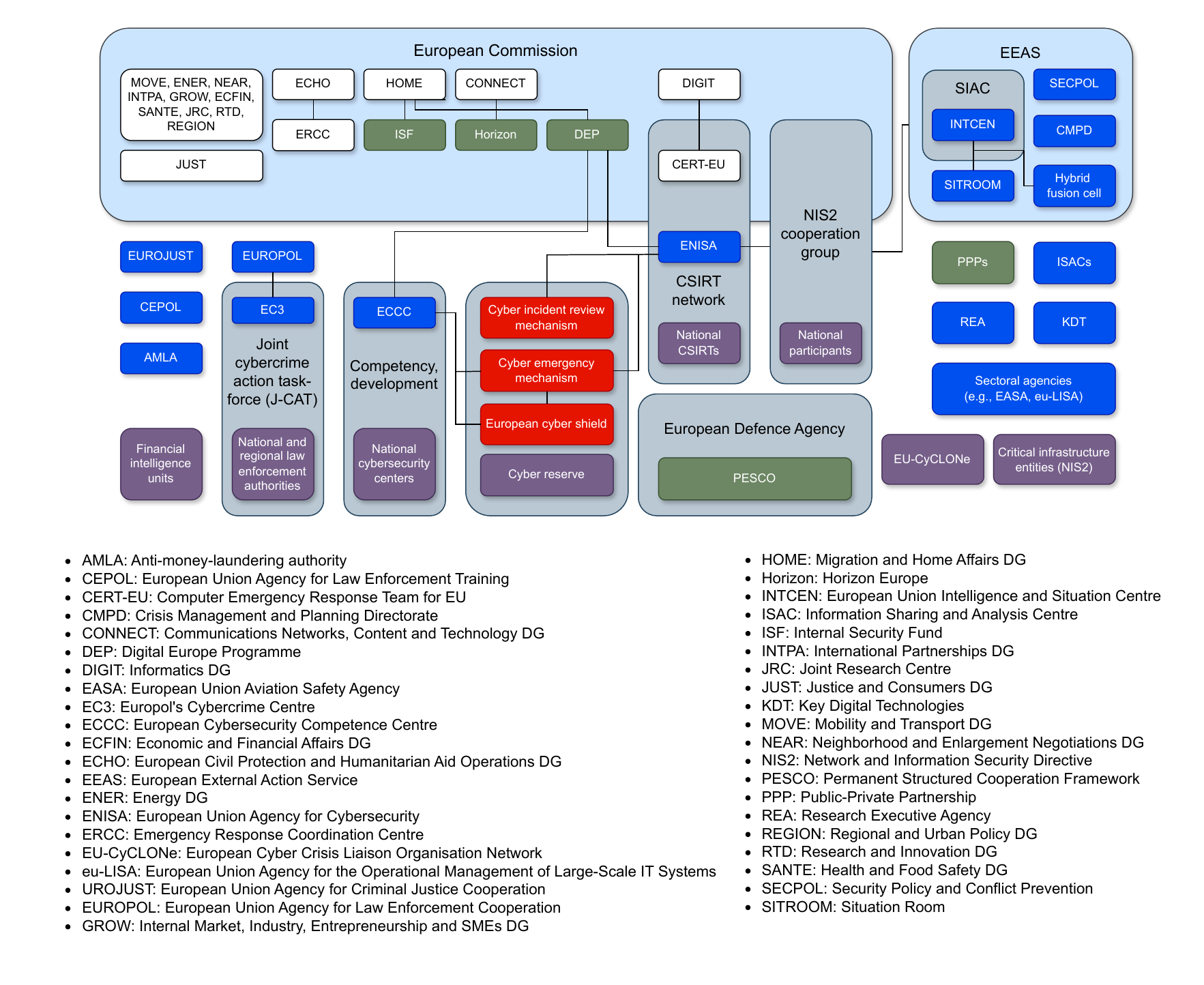}
\caption{Coordination at the EU-level (modified after \cite[p.~14]{ECA24})}
\label{fig: eu}
\end{figure*}

The eight potential policy incoherency is also related to trust. The type of the
incoherency is vertical. During the still ongoing negotiations of the CRA
proposal, considerable criticism was expressed by some member states and
stakeholders about the delivery of information on actively exploited
vulnerabilities to ENISA. The fear was that such a delivery would then spill the
sensitive information to the whole EU and the various actors involved. As with
the cross-border SOCs, national security was part of the fear. The apparent lack
of trust may not be toward ENISA, however, but rather toward the EU-wide CSIRT
network and the other actors involved. While a compromise was found during the
negotiations, information about actively exploited vulnerabilities is still
delivered also to ENISA under most conditions~\cite{Bertuzzi23}. Analogously, it
was also feared that manufacturers would be hesitant to participate because
national CSIRTs are trusted among manufacturers in some member states but not
necessarily in others.

Last but not least, a further potential incoherency is present at the
EU-level. As can be concluded from Fig.~\ref{fig: eu}, the EU-level cyber
security administration is highly complex and the new policies have further
increased the complexity. Bureaucratic inertia thus poses a real possibility of
a permanent institutional incoherency. The issue is also vertical because a
member state may have difficulties at formulating a coherent stance at the
EU-level institutions due to the apparent institutional
fragmentation. Horizontally, it remains somewhat unclear how well the roles and
responsibilities are defined with respect to the various EU-level agencies and
arrangements. Indeed, on one hand, according to evaluations, EU agencies tend to
share cyber security information poorly among themselves~\cite{ECA22}. On the
other hand, trust is a potential issue also at the EU-level, and the actively
exploited vulnerabilities in the CRA proposal again provide a good
example. Because information is delivered to ENISA, it may spill also to EU
agencies who may have an incentive to develop exploits for the
vulnerabilities. The examples include the Europol's cyber crime centre (EC3) and
the EU's intelligence apparatus (INTCEN). Despite these potential trust and
other related issues, none of the cyber security policies considered establish
clear firewalls between different EU agencies and bureaucratic administrative
units. While the free flow of information may improve cyber security, it may
also cause cyber insecurities. Finally, a potential incoherency may be present
with respect to the different funding instruments. The DEP instrument is mostly
about straightforward implementation. Therefore, it remains unclear how well it
aligns with the research and innovation goals pursued in the large-scale Horizon
Europe programme. Furthermore, the civilian-military nexus is again present as
rather similar goals are pursued on the military side through the PESCO
financial instrument.

\section{Conclusion}\label{sec: conclusion}

This paper presented a critical reflection on the EU's new cyber security
policies. Four policies were considered: the CER and NIS2 directives and the CSA
and CRA proposals. The paper's focus was on the potential policy incoherencies
that may arise from the four policies. Several such potential incoherencies were
identified. These include traditional problems, including potential divergence
between the member states, industry sectors, and different technologies,
administrative conflicts, and procedural issues. Trust (or lack thereof) is a
newer potential source of problems. As the policies address pan-European
transmission of sensitive security information, including about exploitable
vulnerabilities, a lack of trust between the member states and EU-level agencies
or between the member states themselves may cause problems in the future. This
point applies also to the threat intelligence information shared through the
cross-border SOCs envisioned. These SOCs also demonstrate a potential problem in
terms of integration, interoperability, and other potential technical
obstacles. In addition, many other policy incoherency problems are present,
including with respect to funding instruments and increased EU-level
bureaucracy, among other things.

Regarding administration at the EU-level, a long-standing criticism has been the
lack of a central agency who would be in charge of the union's strategic and
operational cyber security~\cite{CEPS18, Christou16}. While it remains debatable
whether such a centralized administrative unit is even desirable in the EU, it
can be concluded that the ENISA's increased responsibilities and mandates still
do not amount to a role of a so-called ``cyber tsar''. In contrast,
institutional fragmentation has actually further increased with the
establishment of the EU-CyCLONe network and the NIS cooperation group. Both
likely increase transaction costs. If a serious and large-scale, union-wide
cyber security crisis would emerge, it thus remains unclear who would take the
lead, how the response would be formulated, and how effective it would be. It
may be that the Commission would need to step in.

While the paper's reflection was critical already due to the focus on policy
incoherency, it should be emphasized that the four policies considered likely
significantly bolster the union's cyber security posture even under the presence
of potential problems. In a similar vein, it should be emphasized that the
policies also offer new possibilities for European cyber security companies and
professionals. All clouds have silver linings.

Also some other limitations should be briefly acknowledged. To some extent, the
paper shared the traditional tenet in the policy coherency literature---the
rationalization of policy procedures and institutional arrangements in order to
improve coherency and to maximize synergies between disparate policy goals,
which tend to overlook political and economic changes, trade-offs, and
struggles~\cite{Yunita22}. A more in-depth study would be needed to address
these aspects. For instance, the political aspects would offer a promising path
for further research; it would be interesting to know and to better understand
which member states, companies, organizations, or other stakeholders pursued and
lobbied which policy goals and why. An in-depth study would be also needed to
better understand so-called inter-pillar coherency~\cite{Portela12}. Once cyber
crime, cyber defense, and data protection are also taken into account, it can be
better understood how the EU's grand cyber security setup works and what
potential incoherencies it contains. Here, existing work on earlier EU
policies~\cite{Christou16} hints about rather similar issues discussed in this
paper. As already said, furthermore, the paper was a critical reflection and not
an evaluation. Although \textit{potential} incoherency problems were identified,
it may be that these do not amount to \textit{actual} problems in the
future. Therefore, a thorough policy evaluation is also needed at some point in
time.

\bibliographystyle{splncs03}

\end{document}